\documentclass[useAMS,usenatbib]{mn2e}
\def\nodata{ ~$\cdots$~ }
\def\fm{\mbox{$.\!\!^{\mathrm m}$}}
\def\lesssim{\mathrel{\hbox{\rlap{\hbox{\lower4pt\hbox{$\sim$}}}\hbox{$<$}}}}
\def\gtrsim{\mathrel{\hbox{\rlap{\hbox{\lower4pt\hbox{$\sim$}}}\hbox{$>$}}}}
\usepackage{graphicx}

\begin{document}

\title[$JHK$ Observations of Standard Stars]{$JHK$ Observations of Faint Standard
Stars in the Mauna Kea Near-Infrared Photometric System}
\author[S. K. Leggett et al.]
{S. K. Leggett$^{1,2}$\thanks{email: sleggett@gemini.edu}, 
M. J. Currie$^{2,3}$, W. P. Varricatt$^{4}$, T. G. Hawarden$^{2,5}$,
\and 
A. J. Adamson$^{4}$, J. Buckle$^{2,5}$, 
T. Carroll$^{4}$, J. K. Davies$^{2,5}$, 
C. J. Davis$^{4}$, \and T. H. Kerr$^{4}$, 
O. P. Kuhn$^{2,7}$, 
M. S. Seigar$^{2,8}$, T. Wold$^{4}$.\\
\\
$^{1}$ Gemini Observatory, 670 N. A'ohoku Place, Hilo HI 96720, USA; previously Joint 
Astronomy Centre\\
$^{2}$ Previously affiliated with the Joint Astronomy Centre, 660 N. A'ohoku Place, Hilo HI 96720, USA\\
$^{3}$ Rutherford Appleton Laboratory, Didcot, Oxfordshire OX11 0QX, UK\\
$^{4}$ Joint Astronomy Centre, 660 N. A'ohoku Place, Hilo HI 96720, USA\\
$^{5}$ Astronomy Technology Centre, Blackford Hill, Edinburgh EH9 3HJ, UK\\
$^{6}$ Cavendish Laboratory, Madingley Road, Cambridge CB3 0HE, UK\\
$^{7}$ LBT Observatory, University of Arizona, 933 N Cherry Ave, Tucson, AZ 85721, USA\\
$^{8}$ Department of Physics \& Astronomy, University of California, Irvine, CA 92697-4575, USA}

\date{August 14th 2006}


\maketitle

\label{firstpage}

\begin{abstract}

$JHK$ photometry in the Mauna Kea Observatory (MKO) near-infrared system is presented for 115 stars.  Of these stars, 79 are UK Infrared Telescope (UKIRT) standards from Hawarden et al., and 42 are 
Las Campanas Observatory (LCO, or NICMOS) standards from Persson et al.  The average brightness
of the sample in all three bandpasses is 11.5 magnitudes, with a range between 10 and 15.  The average number of nights each star was observed is 4, and the average 
of the  internal error of the final results is 0$\fm$011.  These 
$JHK$ data agree with those reported by other groups to 0$\fm$02,
for stars in common, which is consistent with the uncertainties.
The measurements are used to derive colour transformations between the MKO $JHK$ photometric system and the UKIRT, LCO and Two Micron All-Sky Survey (2MASS) systems. The 2MASS$-$MKO data scatter by 0$\fm$05 for redder stars,
which is consistent with a dependence on stellar luminosity: the
2MASS $J$ bandpass includes H$_2$O features in dwarfs and the MKO $K$ bandpass includes CO features in giants.
We stress that colour transformations 
derived for stars whose spectra contain only weak features cannot give accurate transformations for objects with strong absorption features within one, but not both, of the filter bandpasses.
We find evidence of systematic  effects at the 0$\fm$02 level in the photometry of stars with
$J<11$ and $H$,$K<10.5$ presented here and in Hawarden et al..  
This is due to an underestimate of the linearity correction for stars observed with the
shortest exposure times; very accurate photometry of stars approaching the saturation limits of infrared detectors 
which are operated in double-read mode is difficult to obtain.
There are indications that four stars in the sample, GSPC S705-D, FS 116 ( B216-b7), FS 144 (Ser-EC84) and
FS 32 (Feige 108), may be variable.  There are 84 stars in the sample presented here that have
$11 < J < 15$ and $10.5 < H$,$K < 15$, are not suspected to be variable, and have
magnitudes with an estimated error $\leq 0\fm$027; 79 of these have an error of  $\leq 0\fm$020. 
These represent the first published high-accuracy $JHK$ stellar photometry in the MKO near-infrared 
photometric system; we recommend these objects be employed as primary standards for that system.
 
\end{abstract}

\begin{keywords}
infrared:stars --  instrumentation: photometers -- 
techniques: photometric -- methods: observational.
\end{keywords}

\section{Introduction}

In the 1990's infrared astronomy changed radically when array cameras started 
to replace single-channel photometers.
Not only did the field of view and spatial
resolution increase markedly, but much fainter limits could be reached.
One repurcussion was that the existing photometric standards were too bright
for the new detectors, and many groups published lists of fainter standards.
These included: \citet{Bo, Ca, Hu, Ps, Ha, Gt}.  


Towards the end of the 1990's and extending into this decade, 
modern infrared sky surveys have come on line.  To maximise the 
output from the surveys requires a 
well-understood photometric system and a large grid of standard stars. 
Calibration of the Two Micron All Sky Survey (2MASS) is described in
\citet{2M} and of the Deep Near-Infrared Survey (DENIS) in \citet{Fo}.
Even deeper surveys have now started at the United Kingdom Infrared
Telescope (UKIRT, the UKIRT Infrared Deep Sky Survey or UKIDSS, see
\citet{He}),
and at the Canada France Hawaii Telescope (CFHT, their WIRCAM instrument,
\citet{Pu}).  


Another area of recent improvement is photometry at 
longer infrared wavelengths.  The UKIRT group presented 
4-5$\mu$m $L^{\prime}M^{\prime}$ standards in \citet{Le}.
Also, Cohen and collaborators have produced a
series of papers extending absolute flux calibrations from the near-infrared
to the mid-infrared, with particular application to mid-infrared satellites.
The first paper and the most recent paper in this series, respectively, are 
\citet{iras} and \citet{Pr}.
Calibration of the recently launched $Spitzer$ infrared space telescope
is described by \citet{Re}.


Despite this progress, a fundamental problem remained for infrared astronomers 
-- there was no single filter set or photometric system.  Not only did this
mean that it was difficult to compare data, but existing filters tended to be
too broad and include poor regions of the atmosphere, leading to additional
noise and thus lower-accuracy data.  This situation was resolved when A. Tokunaga formed
a consortium to purchase a set of well-defined filters that better matched
the atmospheric windows \citep{Si, To, To05}.  These filters have now been widely
adopted; as well as UKIRT (and UKIDSS) the following facilities have purchased
Mauna Kea Observatory (MKO) near-infrared filter sets:
Arizona, Boston, Cornell, Florida, Hawaii, Kyoto, Montreal, Ohio State, Tohoku, 
Tokyo, Virginia and Wyoming Universities; the Anglo-Australian Observatory, CalTech, 
Center for Astrophysics, CFHT, European Southern Observatory, Gemini, 
Grenoble Observatory, the NASA Infrared Telescope Facility, the Isaac Newton Group, 
Keck, Korea Observatory, National Astronomical Observatory of Japan, National Optical
Astronomy Observatory, Nordic Optical Telescope, 
Osservatorio Astrofisico de Arcetri, Rome Observatory, and Telescopio Nazionale Galileo.


In this paper we present a set of standard stars observed at UKIRT using
the UKIRT Fast Track Imager (UFTI) and
$JHK$ filters in the MKO system.
The sample is described in \S 2, the photometric system in \S 3,
the observing and analysis techniques in \S 4 and the results are given in \S 5.
Comparison to published observations of the sample 
are given in \S 6. Our conclusions are given in \S 7.

\begin{figure}
\includegraphics[scale=.45]{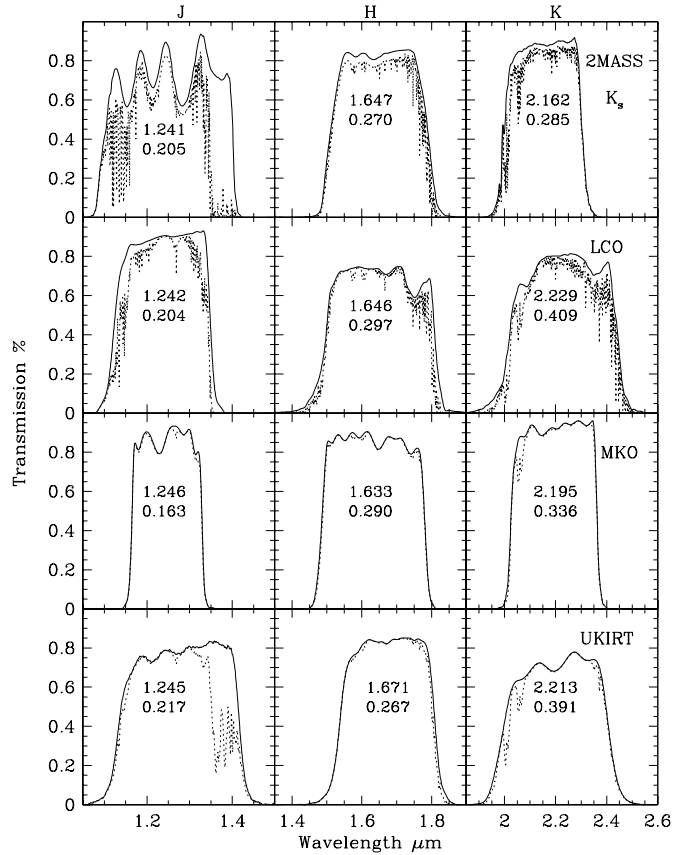}
  \caption{$JHK$ filter transmission profiles, where the dotted line includes typical site-dependent 
atmospheric effects. Central wavelengths and half-power bandwidths are indicated for
filter$+$atmosphere profiles.  See \citet{To05} for a discussion of isophotal and
effective wavelengths for the MKO filter set.}
\label{filt-fig}
\end{figure}

\section[]{The Sample}

The sample consists primarily of UKIRT Faint Standards \citep{Ha} and
Las Campanas (LCO, or NICMOS) standards \citep{Ps}.
Two stars from Hawarden et al., FS 18 and 118, have been removed from this sample due 
to the presence of a nearby star that makes aperture photometry less accurate.  
Three additional unreddened A0 stars, or stars with zero colours, were taken from
\citet{La}.  Such stars are useful for extrapolating magnitudes to other bandpasses. 
The sample is listed in Table 1; coordinates are taken from the Simbad astronomical database,
and proper motions and types from sources provided by the database.
References are given in Table 1.  For stars without spectral types
in the literature we have estimated an approximate type from colour, where possible.

\begin{table*}
 \centering
 \begin{minipage}{300mm}
  \caption{Standard Stars.}
\begin{tabular}{@{}rlrrccccc@{}}
  \hline
   UKIRT &  Name  & \multicolumn{2}{c}{Coordinates (J2000)} & \multicolumn{2}{c}{Proper Motion (mas/yr)} & P. 
M. &Spectral & Type\\
   FS No. &       & RA & Dec &   RA       &         Dec               & Ref. & Type    & Ref.   \\
 \hline
     101 & CMC 400101       &  0:13:43.6 & $+$30:38:00 & $-$5.3 & $-$8.9 & 1 & F0 & 2  \\
 \nodata & BRI B0021$-$0214 &  0:24:24.6 &  $-$1:58:20 & $-$136  & $+$162 & 3 & M9.5V & 4   \\
     102 & GSPC P525-E      &  0:24:28.4  &  $+$7:49:02  & \nodata & \nodata & \nodata & G3  & 5 \\
       1 & G158-100         &  0:33:54.6 & $-$12:07:59 & $+$154  & $-$183 & 6 & DK-G & 7  \\
     103 & GSPC P241-G      &  0:36:29.6 & $+$37:42:54 & \nodata & \nodata & \nodata & K2 & 5  \\
       2 & SA92-342         &  0:55:09.9 & $+$0:43:13  & $-$2.2  & $-$0.2 & 1 & F5 & 8  \\
 \nodata & GSPC S754-C      &  1:03:15.9  & $-$4:20:44    & \nodata & \nodata & \nodata & F/G$^\dag$  & \nodata \\
       3 & Feige 11         &  1:04:21.7 & $+$4:13:37  & $+$12.3 & $-$28.6 & 9 & sdB & 10  \\
     104 & GSPC P194-R      &  1:04:59.6 & $+$41:06:31 & $+$0.4 & $-$4.3 & 1 & A7 & 5   \\
     105 & GSPC P527-F      &  1:19:08.2 & $+$7:34:12  & \nodata & \nodata & \nodata & K1 & 5  \\
     106 & GSPC P152-F      &  1:49:46.9 & $+$48:37:53 & \nodata & \nodata &  \nodata & K4 & 5  \\
     107 & CMC 600954       &  1:54:10.0 & $+$45:50:38 & $-$25.1 & $-$4.0 & 11 & G0 & 2  \\
       5 & Feige 16         &  1:54:34.7 & $-$6:46:00  & \nodata & \nodata & \nodata & A0 & 7  \\
       4 & SA 93-317        &  1:54:37.7 & $+$0:43:00  & $-$9.3  & $-$18.5 & 1 & F5 & 8   \\
       6 & Feige 22         &  2:30:16.6 & $+$5:15:51  & $+$71.2 & $-$24.6 & 9 & DA3 & 12  \\
 \nodata & GSPC P530-D      &  2:33:32.2  & $+$6:25:38    & \nodata & \nodata & \nodata  & G$^\dag$ & \nodata \\
       7 & SA 94-242        &  2:57:21.2 & $+$0:18:39  & \nodata & \nodata & \nodata & A2  & 8 \\
     108 & CMC 502032       &  3:01:09.8 & $+$46:58:48 & $+$1.1 & $-$0.7 & 1 & F8 & 2   \\
 \nodata & TVLM 832-38078   &  3:04:01.8  & $+$00:45:50   & \nodata & \nodata & \nodata & M-V & 13  \\
     109 & LHS 169          &  3:13:24.2 & $+$18:49:38 & $+$1283 & $-$1061 & 3 & esdK7 &  14  \\
 \nodata & GSPC P247-U      &  3:32:03.0  & $+$37:20:41   & \nodata & \nodata & \nodata & G/K$^\dag$ & \nodata \\
     110 & GSPC P533-d      &  3:41:02.4  & $+$6:56:13    & \nodata & \nodata & \nodata & G5  & 5 \\
     111 & CMC 601790       &  3:41:08.6 & $+$33:09:36 & $+$2.5  & $+$3.0 & 11 & G5 & 2 \\
     112 & GSPC S618-D      &  3:47:40.7 & $-$15:13:14 & \nodata & \nodata & \nodata & G0 & 5  \\
      10 & GD 50            &  3:48:50.2 & $-$0:58:31  & \nodata & \nodata & \nodata & DA1 & 12 \\
     113 & GSPC P117-F      &  4:00:14.1 & $+$53:10:39 & \nodata & \nodata & \nodata & K0 & 5   \\
     114 & Melotte 25 LH 214 & 4:19:41.6  & $+$16:45:22   & \nodata & \nodata & \nodata & M7V & 15  \\
     115 & B216-b5          &  4:23:18.2  & $+$26:41:15   & \nodata & \nodata & \nodata &  \nodata & \nodata \\
     116 & B216-b7          &  4:23:50.2  & $+$26:40:07   & \nodata & \nodata & \nodata &  \nodata & \nodata \\
     117 & B216-b9          &  4:23:56.5  & $+$26:36:38   & \nodata & \nodata & \nodata & \nodata & \nodata  \\
      11 & SA 96-83         &  4:52:58.9 & $-$0:14:41  & $+$0.3  & $-$3.2 & 1 & A3 & 8   \\
     119 & HD 289907        &  5:02:57.5 & $-$1:46:43  & $+$1.3 & $-$5.7 & 1 & A2 & 2  \\
 \nodata & GSPC S840-F      &  5:42:32.2  & $+$0:09:04    & \nodata & \nodata & \nodata  & G$^\dag$ & \nodata  \\
      12 & GD 71            &  5:52:27.6 & $+$15:53:13 & $+$92   & $-$189 & 12 & DA1 & 12  \\
      13 & SA 97-249        &  5:57:07.6 & $+$0:01:12  & $+$19.3 & $+$2.4 & 1 & G5V  & 8 \\
     120 & LHS 216          &  6:14:01.2 & $+$15:09:53 & $+$628  & $-$1249 & 3 & sdM2 & 14    \\
 \nodata & GSPC S842-E      &  6:22:43.7  & $-$0:36:30    & \nodata & \nodata &\nodata & G/K$^\dag$ &\nodata  \\
 \nodata & SA 98-653        &  6:52:05.0 & $-$0:18:18  &  $-$0.3 & $-$3.5  & 1 & B9  & 8 \\
     121 & GSPC S772-G      &  6:59:46.8 & $-$4:54:33  & \nodata & \nodata & \nodata & K3 & 5  \\
     122 & GSPC P161-D      &  7:00:52.0  & $+$48:29:24   & \nodata & \nodata & \nodata & G0 & 5  \\
      14 & Rubin 149A       &  7:24:14.0    & $-$0:32:42    & \nodata & \nodata & \nodata & O9-B2p & 7  \\
 \nodata & Rubin 149D       &  7:24:15.4 & $-$0:32:48  & $+$0.70 & $-$8.30 & 1 & A0 & 2   \\
 \nodata & GSPC P309-U      &  7:30:34.6  & $+$29:51:12   & \nodata & \nodata & \nodata  & G$^\dag$ & \nodata \\
 \nodata & GSPC S495-E      &  8:27:12.5  & $-$25:08:01   & \nodata & \nodata & \nodata & G/K$^\dag$ & \nodata  \\
 \nodata & GSPC P545-C      &  8:29:25.2  & $+$5:56:08    & \nodata & \nodata & \nodata & F$^\dag$ & \nodata  \\
 \nodata & LHS 2026         &  8:32:30.5 & $-$1:34:39  & $+$155  & $-$473 & 3 & M6Ve & 16    \\
 \nodata & GSPC S705-D      &  8:36:12.5  & $-$10:13:39   & \nodata & \nodata & \nodata & F/G$^\dag$ & \nodata  \\
      15 & M67-I-48         &  8:51:05.7 & $+$11:43:46 & \nodata & \nodata & \nodata & G5IV-V & 5   \\
     123 & GSPC P486-R      &  8:51:11.8 & $+$11:45:22   & $-$7.7 & $-$5.6 & 9 & B8V & 17   \\
      16 & M67-IV-8         &  8:51:15.1 & $+$11:49:21 & \nodata & \nodata & \nodata & G1V & 5   \\
      17 & M67-IV-2         &  8:51:19.7 & $+$11:52:11 & \nodata & \nodata & \nodata & G4V & 5   \\
     124 & LHS 254          &  8:54:12.3 & $-$8:05:00  & $+$919  & $-$789 & 3 & M6.5V  & 14 \\
     125 & GSPC P259-C      &  9:03:20.6 & $+$34:21:04 & \nodata & \nodata & \nodata & G8 & 5  \\
     126 & GSPC P487-F      &  9:19:18.7 & $+$10:55:54 & \nodata & \nodata & \nodata & K3 & 5  \\
 \nodata & GSPC S852-C      &  9:41:35.9  & $+$0:33:12    & \nodata & \nodata & \nodata & G$^\dag$ & \nodata  \\
 \nodata & GSPC S708-D      &  9:48:56.5  & $-$10:30:32   & \nodata & \nodata & \nodata & K$^\dag$ & \nodata  \\
     127 & GSPC P212-C      & 10:06:29.0  & $+$41:01:27 & \nodata & \nodata & \nodata & F9  & 5 \\
      19 & G 162-66         & 10:33:42.8 & $-$11:41:38 & $-$342  & $-$27.2 & 18  & DA2 & 12  \\
 \nodata & GSPC P550-C      & 10:33:51.9  & $+$4:49:05    & \nodata & \nodata & \nodata & F$^\dag$ & \nodata  \\
     128 & LHS 2347         & 11:05:10.5 & $+$7:06:50  & $-$466  & $-$151 & 19 & M5V  & 16 \\
      20 & G 163-50         & 11:08:00.0 & $-$5:09:26  & $-$63   & $-$440 & 3 & DA3 & 12  \\
\hline
\end{tabular}
\end{minipage}
\end{table*}
\begin{table*}
\begin{minipage}{300mm}
\contcaption{}
\begin{tabular}{@{}rlrrccccc@{}}
  \hline
   UKIRT &  Name  & \multicolumn{2}{c}{Coordinates (J2000)}   & \multicolumn{2}{c}{Proper Motion (mas/yr)} & P. 
M. & Spectral & Type \\
   FS No. &       & RA & Dec &   RA       &         Dec               & Ref. & Type   & Ref.     \\
 \hline
     129 & LHS 2397aAB      & 11:21:49.3 & $-$13:13:08 & $-$508  & $-$80 & 3  & M8V+L7.5 & 20  \\
     130 & GSPC P264-F      & 11:24:55.9 & $+$34:44:39 & \nodata & \nodata & \nodata & K4 & 5  \\
      21 & GD 140           & 11:37:05.1 & $+$29:47:58 & $-$146.9 & $-$5.9 & 9 & DA3  & 12 \\
     131 & GSPC P266-C      & 12:14:25.5  & $+$35:35:55   & \nodata & \nodata & \nodata & F8 & 5  \\
     132 & GSPC S860-D      & 12:21:39.4  & $-$0:07:13    & \nodata & \nodata & \nodata & G1 & 5  \\
      33 & GD 153           & 12:57:02.3 & $+$22:01:53 & $-$33   & $-$206 & 3 & DA1 & 12  \\
     133 & GSPC P172-E      & 13:15:52.8 & $+$46:06:37 & \nodata & \nodata & \nodata & G9 & 5  \\
 \nodata & GSPC S791-C      & 13:17:29.6  & $-$5:32:37    & \nodata & \nodata & \nodata  & F/G$^\dag$ & \nodata \\
      23 & M3-193           & 13:41:43.7 & $+$28:29:51 & $+$3.4  & $+$2.0 & 21 & G8III  & 5 \\
 \nodata & GSPC P133-C      & 13:58:40.3  & $+$52:06:24   & \nodata & \nodata & \nodata  & F$^\dag$ & \nodata \\
 \nodata & GSPC P499-E      & 14:07:34.0  & $+$12:23:51   & \nodata & \nodata & \nodata  & G$^\dag$ & \nodata \\
     134 & LHS 2924         & 14:28:43.3 & $+$33:10:38 & $-$337  & $-$747 & 19 & M9Ve  & 16  \\
     135 & GSPC S867-V      & 14:40:58.0  & $-$0:27:48    & \nodata & \nodata & \nodata & G5  & 5 \\
\nodata  & GSPC P272-D      & 14:58:33.1  & $+$37:08:33   & \nodata & \nodata & \nodata & G$^\dag$ & \nodata  \\
     136 & GSPC S868-G      & 14:59:32.1 & $-$0:06:17  & \nodata & \nodata & \nodata & K2 & 5  \\
 \nodata & TVLM 868-53850   & 15:00:26.4  & $-$0:39:28    & \nodata & \nodata & \nodata & M5Ve  & 22 \\
 \nodata & TVLM 868-110639  & 15:10:17.2  & $-$2:41:07    & \nodata & \nodata & \nodata & M9V & 4  \\
 \nodata & GSPC S870-T      & 15:39:03.6  & $+$0:14:54    & \nodata & \nodata & \nodata  & G$^\dag$ & \nodata  \\
 \nodata & GSPC P177-D      & 15:59:14.0 & $+$47:36:42 & \nodata & \nodata & \nodata & G2V & 23  \\
     137 & GSPC P565-C      & 16:26:42.8  & $+$5:52:20    & \nodata & \nodata & \nodata  & G1 & 5  \\
     138 & GSPC P275-A      & 16:28:06.7 & $+$34:58:48 & $-$11.7 & $+$5.3 & 1 & A1 & 5   \\
 \nodata & GSPC P330-E      & 16:31:33.9 & $+$30:08:47 & \nodata & \nodata & \nodata  & G2V & 23 \\
     139 & GSPC P137-F      & 16:33:53.0 & $+$54:28:22 & \nodata & \nodata  & \nodata  & K1 & 5  \\
      27 & M13-A14          & 16:40:41.3 & $+$36:21:13 & $+$3.9 & $-$1.8 & 24  &  G8IV/V & 5  \\
     140 & GSPC S587-T      & 17:13:22.7 & $-$18:53:34 & \nodata & \nodata & \nodata  & G9  & 5 \\
 \nodata & GSPC P138-C      & 17:13:44.6  & $+$54:33:21   & \nodata & \nodata & \nodata & G$^\dag$ & \nodata  \\
     141 & P489-D           & 17:48:58.9 & $+$23:17:44 & \nodata & \nodata & \nodata & G2 & 5  \\
      35 & GSC 00441-01200  & 18:27:13.5 & $+$4:03:09  & $-$9.0  & $-$4.4 & 18 & K0 & 5   \\
     143 & Ser-EC68         & 18:29:53.9  & $+$1:13:31    & \nodata & \nodata & \nodata & \nodata & \nodata  \\
     144 & Ser-EC84         & 18:29:57.0 & $+$1:12:47  & \nodata & \nodata & \nodata & \nodata & \nodata  \\
 \nodata & GSPC P182-E      & 18:39:33.7  & $+$49:05:38   & \nodata & \nodata & \nodata  & G$^\dag$ & \nodata \\
 \nodata & LDN 547          & 18:51:15.6  & $-$4:16:02    & \nodata & \nodata & \nodata & \nodata & \nodata  \\
     146 & GSPC P280-U      & 18:54:04.0 & $+$37:07:19 & \nodata & \nodata & \nodata & K1 & 5  \\
     147 & GSPC P230-A      & 19:01:55.3 & $+$42:29:19 & $-$1.4 & $-$0.1 & 1 & A0  & 2 \\
 \nodata & GSPC S808-C      & 19:01:55.5  & $-$4:29:12    & \nodata & \nodata & \nodata & G/K$^\dag$  & \nodata  \\
     148 & GSPC S810-A      & 19:41:23.4 & $-$3:50:57  & $-$1.5 & $-$4.5 & 9 & A0 & 25   \\
     149 & GSPC P338-C      & 20:00:39.2 & $+$29:58:38 & $+$4.5 & $-$4.3 & 1 & B7.5V  & 26 \\
     150 & CMC 513807       & 20:36:08.4 & $+$49:38:24 & $+$8.2 & $+$9.3 & 1 & G0 & 2   \\
 \nodata & GSPC S813-D      & 20:41:05.2  & $-$5:03:42    & \nodata & \nodata & \nodata  & G$^\dag$ & \nodata \\
      34 & EG 141           & 20:42:34.8 & $-$20:04:35 & $+$354.6 & $-$97.9 & 9 & DA2.5  & 12  \\
 \nodata & GSPC P576-F      & 20:52:47.4  & $+$6:40:05    & \nodata & \nodata & \nodata & G$^\dag$  & \nodata \\
     151 & GSPC P340-H      & 21:04:14.8 & $+$30:30:21 & \nodata & \nodata & \nodata & G2 & 5  \\
      29 & G 93-48          & 21:52:25.4 & $+$2:23:20  & $+$23.0 & $-$303.0 & 9 & DA3 & 12   \\
 \nodata & BRI B2202-1119   & 22:05:35.7 & $-$11:04:29 & $-$271  & $-$170 & 3 & M5.5V & 4   \\
     152 & GSPC P460-E      & 22:27:16.1 & $+$19:16:59 & \nodata & \nodata & \nodata  & K1  & 5 \\
      30 & SA 114-750       & 22:41:44.7 & $+$1:12:36  & \nodata & \nodata  & \nodata  & B9  & 8 \\
     153 & S820-E           & 23:02:32.1 & $-$3:58:53  & \nodata & \nodata & \nodata  & K2 & 5  \\
      31 & GD 246           & 23:12:23.1 & $+$10:47:04 & $+$127 & $-$11 & 12 & DA1 & 12  \\
      32 & Feige 108        & 23:16:12.4 & $-$1:50:35  & \nodata & \nodata & \nodata & sdB & 27   \\
     154 & GSPC S893-D      & 23:18:10.1  & $+$0:32:57    & \nodata & \nodata & \nodata & G0 & 5  \\
 \nodata & GSPC S677-D      & 23:23:34.5  & $-$15:21:06   & \nodata & \nodata & \nodata  & F$^\dag$  & \nodata \\
 \nodata & GSPC P290-D      & 23:30:33.5  & $+$38:18:57   & \nodata & \nodata & \nodata & G$^\dag$ & \nodata\\
 \nodata & PG 2331$+$055A   & 23:33:44.5  & $+$5:46:41  & \nodata & \nodata & \nodata  & sd & 10 \\
     155 & CMC 516589       & 23:49:47.8 & $+$34:13:05 & \nodata & \nodata & \nodata & K5V & 28   \\

\hline
\end{tabular}
\end{minipage}\\
$^\dag$ Spectral type estimated from $B-V$ and $V-K$ from Lasker et al. 1988 and this work.\\
References: (1) The Tycho Reference Catalogue, Hog et al. 1998 (2) AGK3 Catalogue, Heckmann \& Dieckvoss  1975
(3) Salim \& Gould 2003 (4) Kirkpatrick, Henry \& Simons 1995 (5) Hawarden et al. 2001 (estimated from colour or
colour-magnitude diagrams)
(6)  Harrington \& Dahn 1980 (7) Turnshek et al. 1990 (8) Drilling \& Landolt 1979 (9) The Hipparcos Catalogue,
Perryman et al. 1997 (10) The Palomar-Green Survey, Green, Schmidt \& Liebert 1986 
(11) PPM North Star Catalogue, Roeser \& Bastian 1988 (12) McCook \& Sion 1999 (13) Tinney 1993 (14) Gizis 1997
(15) Reid \& Hawley 1999 (16) Leggett 1992 (17) Pesch 1967 (18) The Second U.S. Naval Observatory CCD 
Astrograph Catalog (UCAC2), Zacharias et al. 2004 (19) Bakos, Sahu \& N\'{e}meth 2002 (20) Freed, Close
\& Siegler 2003 (21) Tucholke, Scholz \& Brosche 1994 (22) Gizis 2002 (23)  Colina \& Bohlin 1997 (24) Kadla 1966
(25) Henry Draper Catalogue, Cannon \& Pickering 1989 (26) Straizys \& Kalytis 1981 
(27) Greenstein \& Sargent 1974 (28) Stephenson 1986.
\\
\end{table*}

\section{Photometric Systems}

\subsection{Filters}

Figure 1 plots the transmission curves of the MKO consortium $JHK$ filters
as well as those of the UKIRT filter set (used by Hawarden et al. 2001), 
the set used by Persson et al. (1998: LCO) and the 2MASS filters.
 Colour transformations between these systems are discussed in \S 6.2.
Profiles are shown with and without atmospheric absorption.
Atmospheric transmissions appropriate for each site were taken from the sources 
given by \citet{St}.
The $J$ and $K$ MKO filters are narrower than the other filter sets (except for 2MASS 
$K_{\rm s}$) to avoid the poor regions of the atmospheric windows. The 2MASS and
UKIRT  $J$-band filters
are particularly prone to atmospheric absorption and therefore noise.
For further discussion, see \citet{Si} and \citet{To}.

\subsection{Optical Elements}

As summarised in \citet{Ha, Le, St},
the commonly used near-infrared optical elements have transmission or absorption 
wavelength responses that are flat across the $JHK$ filter bandpasses.  Hence instrument
and telescope optics typically do not significantly affect the
photometric system. \citet{Ha} found that even changes
in UKIRT's optical/infrared dichroic coating did not affect the $J$ bandpass.
\citet{St} did find that, for the very structured spectra of late L and T dwarfs, the
variations between detector quantum efficiency and anti-reflection coatings can 
affect measured magnitudes by up to 1\%.  
However for stars with a more typical spectral energy distribution, such as those in this sample, 
these effects are negligible.
Hence the results presented here are appropriate for any camera equipped with the
$JHK$ MKO filters and a detector with a reasonably flat
sensitivity curve (such as UKIRT's Wide Field Camera WFCAM -- but see further discussion in \S 6.2).

\section{Techniques}

\subsection{Observations}

All observations presented here were made at UKIRT using the 
UKIRT Fast Track Imager, UFTI \citep{ufti}.  UFTI contains a HAWAII-I
HgCdTe detector, with 1024$\times$1024 0.091 arcsecond pixels.
For standard star observations a 512$\times$512 subarray readout was used,
allowing minimum exposure times of 1 second. UFTI was
outfitted with MKO filters from its commissioning in 1998 October.

The data presented here were taken over 21 photometric nights between 2001 January 23 and 
2004 December 29.
10 to 50 standards were observed on whole or partial nights.
One to three would be repeated to measure the extinction on each night and
the stars would be observed between one and $\sim$two airmasses.
Exposure times ranged from 1 second to 60 seconds,
but were usually around 8 seconds at $J$ and 4 seconds at $H$ and $K$.
On nights of very poor or very good seeing the exposure times
would be adjusted to keep the peak data counts on the star between
500 and 4000 data numbers, as far as possible.
Typically a five-position jitter pattern was used with 10-arcsecond
offsets.  Nine-position jitters were used for stars fainter than 15th magnitude.

\subsection{Data Reduction}

The data were reduced using the ORAC-DR pipeline (Cavanagh et al. 2003).
Details of the reduction recipes (JITTER\_SELF\_FLAT\_APHOT, BRIGHT\_POINT\_SOURCE\_APHOT)
are presented in the ORAC-DR Imaging Data Reduction User
Guide\footnote{http://www.starlink.ac.uk/star/docs/sun232.htx/sun232.html}).
ORAC-DR uses various Starlink  software packages
to reduce astronomical data. 

The raw image counts are first corrected for non-linearity using the empirical
correction determined at the observatory in 2000 October:
$$ {\rm true} = {\rm measured}/(1 + (6.1.10^{-6}*{\rm measured})) $$
Note that the measured counts are {\it larger} than the corrected counts.
This relationship holds for measured counts fewer than 10000 data numbers (DN) --
above this limit the detector goes into hard saturation where measured counts will be
much {\it smaller} than predicted.
The non-linearity correction is small -- around 2\% at the typical value of
4000 counts.  We discarded data where any pixel was higher than 10000~DN. 

The raw images, and the engineering data used to derive the non-linearity
correction, were taken in non-destructive read mode, where the detector
is reset, read out, exposed and read out again, and the counts in the
stored image are the difference between the two reads.  The linearity
correction we have applied to the images is not strictly correct, as a
correction should be applied to each individual read and not the difference
\citep{Ps,Vc}. However in \S 6.3 we show that the error is insignificant except
for the brightest stars in the sample which were observed with
exposure times similar to the array readout time; for these stars the effect is 
small but significant. Note that UFTI's non-linearity correction
goes in the opposite sense to the more typical well-filling saturation
behaviour, where lower, not higher, counts would be expected as signal 
increases.  The  somewhat unusual UFTI controller may be the 
cause of this linearity behaviour.  The controller is derived
from the CIRSI controller \citep{cirsi} which in turn is based on
an LSR-Astrocam (later PerkinElmer Life Science) 4100 CCD controller.
During the read of each pixel, a fast video switch in the interface switches 
between the real output of the array and a dummy level,
to remove drifts.

\begin{table*}
 \centering
  \caption{MKO $JHK$ Extinction (Magnitudes/Airmass).}
\begin{tabular}{@{}crrrrrr@{}}
  \hline
   Filter &  Min.  & Max. & Mean & Std.  & \multicolumn{2}{c}{Calculated}\\
          &        &      &      & Devn. & \multicolumn{2}{c}{for pwv:}\\
          &        &      &       &           &  1mm  & 4mm  \\ 
 \hline
      $J$  & 0.012 & 0.090 & 0.047 & 0.024 & 0.006 & 0.014 \\
      $H$  & 0.000 & 0.072 & 0.029 & 0.020 & 0.005 & 0.015 \\
      $K$  & 0.000 & 0.124 & 0.052 & 0.028 & 0.017 & 0.019 \\

\hline
\end{tabular}
\end{table*}

After correcting for non-linearity, each frame is bad-pixel 
masked, dark-subtracted and flat-fielded. 
Dark frames are taken with every star.  Any darks that suffered from latent
images are rejected. Flat fields are created from median-filtered and 
object-masked jittered images of the fainter standards. Object masking
detects objects with 12 connected pixels at 1~$\sigma$ above the sky level.
The locations, shapes, orientations and sizes are used to make a mask which 
is applied to the dark-subtracted frames to generate the flat field.
Jittered frames of fainter stars are used to generate the flat field so 
that the background level is dominated by sky noise and not read noise.
For UFTI, images taken with $J$ exposure times $\geq$ 20 seconds, and 
$H$ and $K$ exposure times $\geq$ 5 seconds, are background-noise-limited.  
For brighter standards that were not background-noise-limited,
the closest-in-time flat field, generated with a faint standard, was used.

Finally, a mosaic is created from the flat-fielded jitter set of five or nine pointings.  
Aperture photometry is carried out on each source using a 7-arcsecond diameter aperture, 
with sky annuli from 10.5 to 17.5 arcseconds.  This is large enough to include all the light from 
the star even on nights of mediocre seeing, without compromising 
signal-to-noise. On one night with poor seeing (20020217) an 8-arcsecond aperture had to be used.

ORAC-DR reports photometric errors on each individual flat-fielded 
frame, as well as on the mosaic.  For this work we have conservatively adopted the
uncertainty to be the standard deviation in the mean of the measurements 
of a jitter set. This standard deviation is 1--2\%, compared to the 
millimag-level error calculated by ORAC-DR from the sky variance of the mosaic.  

Zeropoint magnitudes are calculated for each measurement by subtracting the
instrumental magnitude from the catalogue magnitude.  These zeropoints
are plotted against airmass for each filter for each night.
A linear extinction curve is fitted to the
data using the extinction star(s), and each catalogue magnitude adjusted 
so that the star would agree with the average zeropoint at an airmass of 
one, as defined by that night's sample.  In this way we revised the standard
stars' catalogue of $JHK$ magnitudes after every observing run, such that
the group became more self-consistent as time went on.  The initial standard star
catalogue consisted of UKIRT Faint Standard values transformed to the
MKO system using the preliminary colour terms given in \citet{Ha}.

For this paper we re-reduced the data from all 21 nights using the most
recent version of the catalogue, to ensure that the entire set was
handled consistently.  Linear extinction curves were fitted for each night
in each filter using the
entire night's data (i.e. not just the extinction stars). The scatter around 
this curve was $0\fm02$ or less.  A single $J$, $H$ or $K$ value was calculated for 
each star for each night and these values then averaged over all nights.
The results are given in the next section.

\section{Results}

\subsection{Extinction}

Table 2 lists our measured extinction values at $JHK$, determined by linear regression
to the [airmass, zeropoint] values (for airmass ranging from 1 to   
around 2), for the sample of 21 nights. The uncertainty in the
nightly extinction value is $\sim 0.01$ magnitudes/airmass. Water vapour measurements over 
these nights ranged from a 225 GHz $\tau$ of 0.04 to 0.27 (average 0.10), corresponding to 
a precipitable water vapour (pwv) of between 1 and 5 mm (average 2 mm).  We find no correlation
between $\tau$ and extinction,  as expected for these filters which are well matched to
the atmospheric windows. 
\citet{To} calculate the extinction due to water vapour for the MKO filters,
for 1 to 4 mm of water vapour, and these values are given in Table 2.
It can be seen that the measured values are higher, which would be expected as extinction
due to scattering is not included in the calculated values. \citet{Tug}, for example,
have shown that aerosol scattering is significant in the far red.  The difference between
the measured and predicted extinction values in Table 2 indicate that scattering can contribute
between 0 and $\sim$0.1 magnitudes/airmass to the extinction at $JHK$, on the summit of Mauna Kea.  
For accurate photometry of targets with airmass differing significantly from their
calibrators, the extinctions must be determined on each night.

\subsection{$JHK$ magnitudes}

\begin{table*}
 \centering
 \begin{minipage}{300mm}
  \caption{Measured MKO $JHK$ magnitudes.}
\begin{tabular}{@{}rlrrcrrcrrcc@{}}
  \hline
   UKIRT &  Name  & \multicolumn{3}{c}{$J_{MKO}$} & \multicolumn{3}{c}{$H_{MKO}$} & 
\multicolumn{3}{c}{$K_{MKO}$} & Note\\
   FS No. &       & mag & Std.Dev. & Nts. & mag & Std.Dev. & Nts.
& mag & Std.Dev. & Nts.  &    \\
 \hline
     101 & CMC 400101       &  10.540 & 0.005 & 4 & 10.405 & 0.012 & 4 & 10.366 & 0.016 & 5 & 1 \\
 \nodata & BRI B0021$-$0214 &  11.754 & 0.007 & 4 & 11.082 & 0.011 & 6 & 10.500 & 0.008 & 6 & \\
     102 & GSPC P525-E      &  11.583 & 0.013 & 3 & 11.273 & 0.008 & 3 & 11.200 & 0.012 & 3 & \\
       1 & G158-100         &  13.427 & 0.006 & 3 & 13.059 & 0.011 & 3 & 12.984 & 0.019 & 3 & \\
     103 & GSPC P241-G      &  12.342 & 0.011 & 4 & 11.840 & 0.008 & 6 & 11.724 & 0.012 & 3 & \\
       2 & SA92-342         &  10.699 & 0.005 & 4 & 10.495 & 0.009 & 4 & 10.470 & 0.009 & 4 & 1 \\
 \nodata & GSPC S754-C      &  11.008 & 0.009 & 4 & 10.721 & 0.008 & 4 & 10.669 & 0.007 & 4 & \\
       3 & Feige 11         &  12.649 & 0.009 & 5 & 12.739 & 0.011 & 4 & 12.840 & 0.012 & 4 & \\
     104 & GSPC P194-R      &  10.503 & 0.008 & 6 & 10.414 & 0.013 & 5 & 10.395 & 0.013 & 4 & 1 \\
     105 & GSPC P527-F      &  11.526 & 0.012 & 5 & 11.070 & 0.010 & 5 & 10.961 & 0.007 & 5 & \\
     106 & GSPC P152-F      &  12.466 & 0.009 & 3 & 11.877 & 0.013 & 3 & 11.752 & 0.010 & 3 & \\
     107 & CMC 600954       &  10.466 & 0.006 & 4 & 10.256 & 0.011 & 5 & 10.224 & 0.010 & 4 & 1 \\
       5 & Feige 16         &  12.359 & 0.017 & 2 & 12.336 & 0.011 & 2 & 12.349 & 0.024 & 3 & \\
       4 & SA 93-317        &  10.538 & 0.011 & 5 & 10.304 & 0.013 & 5 & 10.266 & 0.010 & 5 & 1 \\
       6 & Feige 22         &  13.271 & 0.009 & 4 & 13.321 & 0.004 & 4 & 13.404 & 0.009 & 3 & \\
 \nodata & GSPC P530-D      &  11.267 & 0.008 & 2 & 10.936 & 0.007 & 3 & 10.878 & 0.017 & 3 & \\
       7 & SA 94-242        &  11.076 & 0.010 & 3 & 10.961 & 0.009 & 3 & 10.933 & 0.006 & 4 & \\
     108 & CMC 502032       &  10.056 & 0.014 & 3 &  9.765 & 0.007 & 4 &  9.713 & 0.011 & 4 & 1 \\
 \nodata & TVLM 832-38078   &  11.755 & 0.013 & 4 & 11.221 & 0.011 & 5 & 10.844 & 0.012 & 4 & \\
     109 & LHS 169          &  11.435 & 0.015 & 4 & 10.983 & 0.011 & 4 & 10.813 & 0.010 & 4 & \\
 \nodata & GSPC P247-U      &  11.907 & 0.009 & 4 & 11.606 & 0.007 & 3 & 11.508 & 0.013 & 4 & \\
     110 & GSPC P533-d      &  11.703 & 0.011 & 4 & 11.406 & 0.004 & 4 & 11.321 & 0.007 & 4 & \\
     111 & CMC 601790       &  10.618 & 0.013 & 4 & 10.361 & 0.004 & 3 & 10.275 & 0.016 & 4 & 1 \\
     112 & GSPC S618-D      &  11.182 & 0.010 & 3 & 10.919 & 0.019 & 3 & 10.855 & 0.008 & 2 & \\
      10 & GD 50            &  14.802 & 0.017 & 5 & 14.878 & 0.013 & 6 & 14.990 & 0.015 & 4 & \\
     113 & GSPC P117-F      &  12.915 & 0.009 & 3 & 12.559 & 0.010 & 2 & 12.443 & 0.010 & 3 & \\
     114 & Melotte 25 LH 214 & 14.373 & 0.009 & 4 & 13.876 & 0.010 & 3 & 13.434 & 0.008 & 2 & \\
     115 & B216-b5          &  12.501 & 0.006 & 3 & 10.976 & 0.010 & 3 & 10.073 & 0.013 & 4 & 2 \\
     116 & B216-b7          &  12.775 & 0.008 & 3 & 11.532 & 0.004 & 4 & 10.928 & 0.007 & 4 & 3 \\
     117 & B216-b9          &  11.495 & 0.006 & 3 & 10.560 & 0.008 & 4 & 10.045 & 0.006 & 4 & 2 \\
      11 & SA 96-83         &  11.329 & 0.005 & 3 & 11.264 & 0.013 & 3 & 11.241 & 0.008 & 3 & \\
     119 & HD 289907        &   9.857 & 0.007 & 3 &  9.818 & 0.010 & 3 &  9.810 & 0.010 & 3 & 1 \\
 \nodata & GSPC S840-F      &  11.365 & 0.016 & 3 & 11.099 & 0.013 & 3 & 11.019 & 0.009 & 3 & \\
      12 & GD 71            &  13.710 & 0.009 & 4 & 13.805 & 0.018 & 4 & 13.899 & 0.011 & 3 & \\
      13 & SA 97-249        &  10.471 & 0.009 & 4 & 10.176 & 0.007 & 4 & 10.126 & 0.009 & 4 & 1 \\
     120 & LHS 216          &  11.299 & 0.015 & 3 & 10.855 & 0.011 & 4 & 10.603 & 0.011 & 4 & \\
 \nodata & GSPC S842-E      &  11.660 & 0.020 & 3 & 11.327 & 0.013 & 3 & 11.233 & 0.014 & 3 & \\
 \nodata & SA 98-653        &   9.428 & 0.014 & 5 &  9.422 & 0.016 & 5 &  9.443 & 0.010 & 4 & 1 \\
     121 & GSPC S772-G      &  11.984 & 0.015 & 3 & 11.436 & 0.007 & 3 & 11.307 & 0.014 & 3 & \\
     122 & GSPC P161-D      &  11.668 & 0.008 & 4 & 11.393 & 0.006 & 4 & 11.347 & 0.005 & 4 & \\
      14 & Rubin 149A       &  14.128 & 0.006 & 4 & 14.164 & 0.017 & 5 & 14.214 & 0.016 & 4 & \\
 \nodata & Rubin 149D       &  11.444 & 0.007 & 3 & 11.438 & 0.010 & 3 & 11.459 & 0.010 & 3 & \\
 \nodata & GSPC P309-U      &  11.841 & 0.012 & 5 & 11.507 & 0.010 & 4 & 11.449 & 0.007 & 5 & \\
 \nodata & GSPC S495-E      &  11.483 & 0.014 & 3 & 11.031 & 0.014 & 3 & 10.939 & 0.009 & 2 & \\
 \nodata & GSPC P545-C      &  11.841 & 0.018 & 5 & 11.585 & 0.010 & 4 & 11.549 & 0.010 & 5 & \\
 \nodata & LHS 2026         &  11.990 & 0.010 & 3 & 11.482 & 0.005 & 3 & 11.075 & 0.014 & 3 & \\
 \nodata & GSPC S705-D      &  12.353 & 0.055 & 5 & 12.047 & 0.042 & 5 & 12.027 & 0.048 & 5 & 4 \\
      15 & M67-I-48         &  12.722 & 0.008 & 4 & 12.423 & 0.014 & 5 & 12.359 & 0.013 & 4 & \\
     123 & GSPC P486-R      &  10.126 & 0.012 & 4 & 10.158 & 0.010 & 4 & 10.203 & 0.007 & 4 & 1 \\
      16 & M67-IV-8         &  12.968 & 0.014 & 5 & 12.702 & 0.009 & 5 & 12.650 & 0.015 & 6 & \\
      17 & M67-IV-2         &  12.668 & 0.017 & 3 & 12.363 & 0.010 & 4 & 12.290 & 0.011 & 4 & \\
     124 & LHS 254          &  11.467 & 0.018 & 5 & 11.085 & 0.007 & 4 & 10.727 & 0.009 & 3 & \\
     125 & GSPC P259-C      &  10.797 & 0.006 & 4 & 10.423 & 0.011 & 4 & 10.355 & 0.007 & 4 & 1 \\
     126 & GSPC P487-F      &  12.304 & 0.011 & 3 & 11.772 & 0.007 & 3 & 11.636 & 0.010 & 3 & \\
 \nodata & GSPC S852-C      &  11.288 & 0.019 & 3 & 11.004 & 0.019 & 3 & 10.960 & 0.009 & 2 & \\
 \nodata & GSPC S708-D      &  11.034 & 0.010 & 4 & 10.736 & 0.010 & 4 & 10.671 & 0.013 & 4 & \\
     127 & GSPC P212-C      &  11.969 & 0.007 & 4 & 11.727 & 0.008 & 5 & 11.685 & 0.011 & 5 & \\
      19 & G 162-66         &  13.625 & 0.015 & 6 & 13.691 & 0.012 & 7 & 13.789 & 0.017 & 6 & \\ 
 \nodata & GSPC P550-C      & 12.293 & 0.014 & 3 & 12.083 & 0.004 & 3 & 12.052 & 0.008 & 3 & \\
     128 & LHS 2347         & 12.985 & 0.008 & 3 & 12.421 & 0.005 & 3 & 12.046 & 0.017 & 3 & \\
      20 & G 163-50         & 13.427 & 0.010 & 4 & 13.457 & 0.005 & 5 & 13.509 & 0.006 & 3 & \\
\hline
\end{tabular}
\end{minipage}
\end{table*}
\begin{table*}
\begin{minipage}{300mm}
\contcaption{}
\begin{tabular}{@{}rlrrcrrcrrcc@{}}
  \hline
   UKIRT &  Name  & \multicolumn{3}{c}{$J_{MKO}$} & \multicolumn{3}{c}{$H_{MKO}$} & 
\multicolumn{3}{c}{$K_{MKO}$} & Note\\
   FS No. &       & mag & Std.Dev. & Nts. & mag & Std.Dev. & Nts.
& mag & Std.Dev. & Nts. &    \\
 \hline
     129 & LHS 2397aAB      & 11.792 & 0.005 & 5 & 11.173 & 0.010 & 5 & 10.639 & 0.009 & 5 & \\
     130 & GSPC P264-F      & 12.980 & 0.008 & 2 & 12.402 & 0.005 & 3 & 12.262 & 0.011 & 3 & \\
      21 & GD 140           & 13.021 & 0.007 & 3 & 13.075 & 0.011 & 3 & 13.167 & 0.017 & 3 & \\
     131 & GSPC P266-C      & 11.617 & 0.009 & 3 & 11.362 & 0.010 & 3 & 11.324 & 0.012 & 4 &  \\
     132 & GSPC S860-D      & 12.159 & 0.012 & 4 & 11.879 & 0.015 & 4 & 11.835 & 0.012 & 4 & \\
      33 & GD 153           & 14.085 & 0.007 & 3 & 14.165 & 0.010 & 3 & 14.296 & 0.011 & 2 & \\
     133 & GSPC P172-E      & 12.330 & 0.011 & 3 & 11.971 & 0.006 & 3 & 11.909 & 0.008 & 3 & \\
 \nodata & GSPC S791-C      & 11.605 & 0.011 & 3 & 11.283 & 0.014 & 3 & 11.227 & 0.015 & 4 & \\
      23 & M3-193           & 12.990 & 0.006 & 3 & 12.488 & 0.004 & 3 & 12.397 & 0.009 & 3 & \\
 \nodata & GSPC P133-C      & 11.113 & 0.024 & 4 & 10.876 & 0.006 & 4 & 10.832 & 0.009 & 5 & \\
 \nodata & GSPC P499-E      & 11.893 & 0.008 & 3 & 11.560 & 0.031 & 3 & 11.528 & 0.008 & 2 & \\
     134 & LHS 2924         & 11.885 & 0.013 & 4 & 11.251 & 0.010 & 5 & 10.721 & 0.013 & 5 & \\
     135 & GSPC S867-V      & 11.965 & 0.010 & 2 & 11.669 & 0.007 & 3 & 11.604 & 0.007 & 3 & \\
\nodata  & GSPC P272-D      & 11.601 & 0.008 & 4 & 11.267 & 0.011 & 4 & 11.212 & 0.008 & 4 & \\
     136 & GSPC S868-G      & 12.531 & 0.012 & 4 & 12.014 & 0.009 & 5 & 11.893 & 0.012 & 5 & \\
 \nodata & TVLM 868-53850   & 11.517 & 0.009 & 3 & 10.988 & 0.010 & 3 & 10.617 & 0.014 & 3 & \\
 \nodata & TVLM 868-110639  & 12.530 & 0.020 & 4 & 11.875 & 0.009 & 4 & 11.306 & 0.015 & 2 & \\
 \nodata & GSPC S870-T      & 10.862 & 0.005 & 3 & 10.667 & 0.012 & 3 & 10.632 & 0.012 & 3 & 5 \\
 \nodata & GSPC P177-D      & 12.212 & 0.024 & 4 & 11.920 & 0.006 & 4 & 11.865 & 0.010 & 3 & \\
     137 & GSPC P565-C      & 12.140 & 0.017 & 3 & 11.893 & 0.006 & 3 & 11.838 & 0.009 & 3 & \\
     138 & GSPC P275-A      & 10.391 & 0.015 & 4 & 10.391 & 0.014 & 4 & 10.416 & 0.008 & 3 & 1 \\
 \nodata & GSPC P330-E      & 11.772 & 0.017 & 5 & 11.455 & 0.017 & 5 & 11.419 & 0.011 & 4 & \\
     139 & GSPC P137-F      & 12.671 & 0.023 & 5 & 12.228 & 0.015 & 5 & 12.126 & 0.010 & 5 & \\
      27 & M13-A14          & 13.470 & 0.011 & 4 & 13.199 & 0.005 & 3 & 13.135 & 0.008 & 4 & \\
     140 & GSPC S587-T      & 10.775 & 0.014 & 4 & 10.430 & 0.021 & 4 & 10.369 & 0.008 & 3 & 1 \\
 \nodata & GSPC P138-C      & 11.327 & 0.027 & 2 & 11.124 & 0.007 & 2 & 11.098 & 0.019 & 3 & \\
     141 & P489-D           & 11.152 & 0.015 & 3 & 10.853 & 0.009 & 3 & 10.785 & 0.011 & 3 & \\
      35 & GSC 00441-01200  & 12.188 & 0.015 & 5 & 11.835 & 0.010 & 5 & 11.741 & 0.010 & 5 & \\
     143 & Ser-EC68         & 16.495 & 0.030 & 3 & 14.269 & 0.010 & 2 & 12.923 & 0.011 & 3 & \\
     144 & Ser-EC84         & 14.997 & 0.030 & 3 & 12.555 & 0.013 & 3 & 11.009 & 0.012 & 3 & 4 \\
 \nodata & GSPC P182-E      & 12.081 & 0.006 & 4 & 11.779 & 0.013 & 3 & 11.713 & 0.011 & 3 & \\
 \nodata & LDN 547          & 11.753 & 0.013 & 4 &  9.890 & 0.010 & 3 &  8.889 & 0.005 & 4 & 6 \\
     146 & GSPC P280-U      & 10.708 & 0.008 & 4 & 10.209 & 0.006 & 4 & 10.120 & 0.012 & 4 & 1 \\
     147 & GSPC P230-A      &  9.868 & 0.011 & 4 &  9.839 & 0.004 & 5 &  9.834 & 0.009 & 5 & 1 \\
 \nodata & GSPC S808-C      & 10.925 & 0.013 & 3 & 10.628 & 0.014 & 3 & 10.543 & 0.009 & 3 & 5 \\
     148 & GSPC S810-A      &  9.437 & 0.010 & 5 &  9.423 & 0.008 & 5 &  9.438 & 0.005 & 5 & 1 \\
     149 & GSPC P338-C      & 10.073 & 0.012 & 3 & 10.061 & 0.018 & 5 & 10.061 & 0.015 & 5 & 1 \\
     150 & CMC 513807       & 10.133 & 0.006 & 3 &  9.985 & 0.007 & 3 &  9.941 & 0.011 & 3 & 1 \\
 \nodata & GSPC S813-D      & 11.434 & 0.014 & 4 & 11.118 & 0.009 & 4 & 11.053 & 0.012 & 4 & \\
      34 & EG 141           & 12.883 & 0.009 & 4 & 12.930 & 0.013 & 4 & 13.000 & 0.016 & 2 & \\
 \nodata & GSPC P576-F      & 12.215 & 0.011 & 3 & 11.924 & 0.007 & 3 & 11.854 & 0.015 & 3 & \\
     151 & GSPC P340-H      & 12.211 & 0.014 & 3 & 11.946 & 0.008 & 3 & 11.865 & 0.014 & 3 & \\
      29 & G 93-48          & 13.215 & 0.012 & 4 & 13.255 & 0.010 & 4 & 13.330 & 0.008 & 4 & \\
 \nodata & BRI B2202-1119   & 11.595 & 0.007 & 3 & 11.085 & 0.013 & 4 & 10.708 & 0.010 & 3 & \\
     152 & GSPC P460-E      & 11.648 & 0.013 & 4 & 11.130 & 0.009 & 4 & 11.048 & 0.007 & 4 & \\
      30 & SA 114-750       & 11.936 & 0.005 & 3 & 11.975 & 0.014 & 4 & 12.013 & 0.012 & 4 & \\
     153 & S820-E           & 11.603 & 0.013 & 4 & 11.029 & 0.012 & 4 & 10.890 & 0.009 & 4 & \\
      31 & GD 246           & 13.845 & 0.017 & 3 & 13.961 & 0.009 & 4 & 14.064 & 0.014 & 4 & \\
      32 & Feige 108        & 13.570 & 0.007 & 3 & 13.663 & 0.010 & 3 & 13.746 & 0.008 & 2 & 4 \\
     154 & GSPC S893-D      & 11.373 & 0.008 & 3 & 11.098 & 0.012 & 4 & 11.050 & 0.008 & 3 & \\
 \nodata & GSPC S677-D      & 11.811 & 0.009 & 6 & 11.559 & 0.012 & 5 & 11.537 & 0.014 & 6 & \\
 \nodata & GSPC P290-D      & 11.617 & 0.009 & 3 & 11.331 & 0.009 & 3 & 11.255 & 0.016 & 3 & \\
 \nodata & PG 2331$+$055A   & 15.333 & 0.030 & 4 & 15.381 & 0.031 & 5 & 15.404 & 0.040 & 4 & \\
     155 & CMC 516589       &  9.964 & 0.009 & 4 &  9.472 & 0.016 & 5 &  9.383 & 0.013 & 6 & 1 \\

\hline
\end{tabular}
\end{minipage}\\
(1) $JHK$ may be too bright by $\lesssim$0$\fm$02 due to non-linearity underestimate\\
(2) $K$ may be too bright by $\lesssim$0$\fm$02 due to non-linearity underestimate\\
(3) May be variable at $J$ by $\sim 0\fm1$ over a period of years\\
(4) May be variable at $JHK$ by $0\fm05$--$0\fm10$ over a period of months to years\\
(5) $J$ may be too bright by $\lesssim$0$\fm$02 due to non-linearity underestimate\\
(6) $HK$ may be too bright by $\lesssim$0$\fm$02 due to non-linearity underestimate\\

\end{table*}

Table 3 gives our  weighted mean $JHK$ magnitudes, the estimated error of that mean, 
and the number of nights the star was observed in that filter. The estimated error 
is the larger of (1) the standard deviation of the mean, or
(2) $\sqrt(1/\sum(1/\sigma^2_{\rm night}))$.
Individual night's magnitudes have been
omitted if they deviated by more than  3~$\sigma$ (or typically $> 0\fm03$) from the mean,
as defined by the value of $\sigma$ after the omission.  
As the MKO filter set has been widely adopted 
because of its design advantages, and differs significantly from most earlier filter
sets, including the UKIRT filters used by Hawarden et al (2001), we recommend that the 
present results are used instead of
those of Hawarden et al., despite the latter's superior internal accuracy (0$\fm$005 cf. 0$\fm$011).

\begin{figure}
\includegraphics[scale=.45]{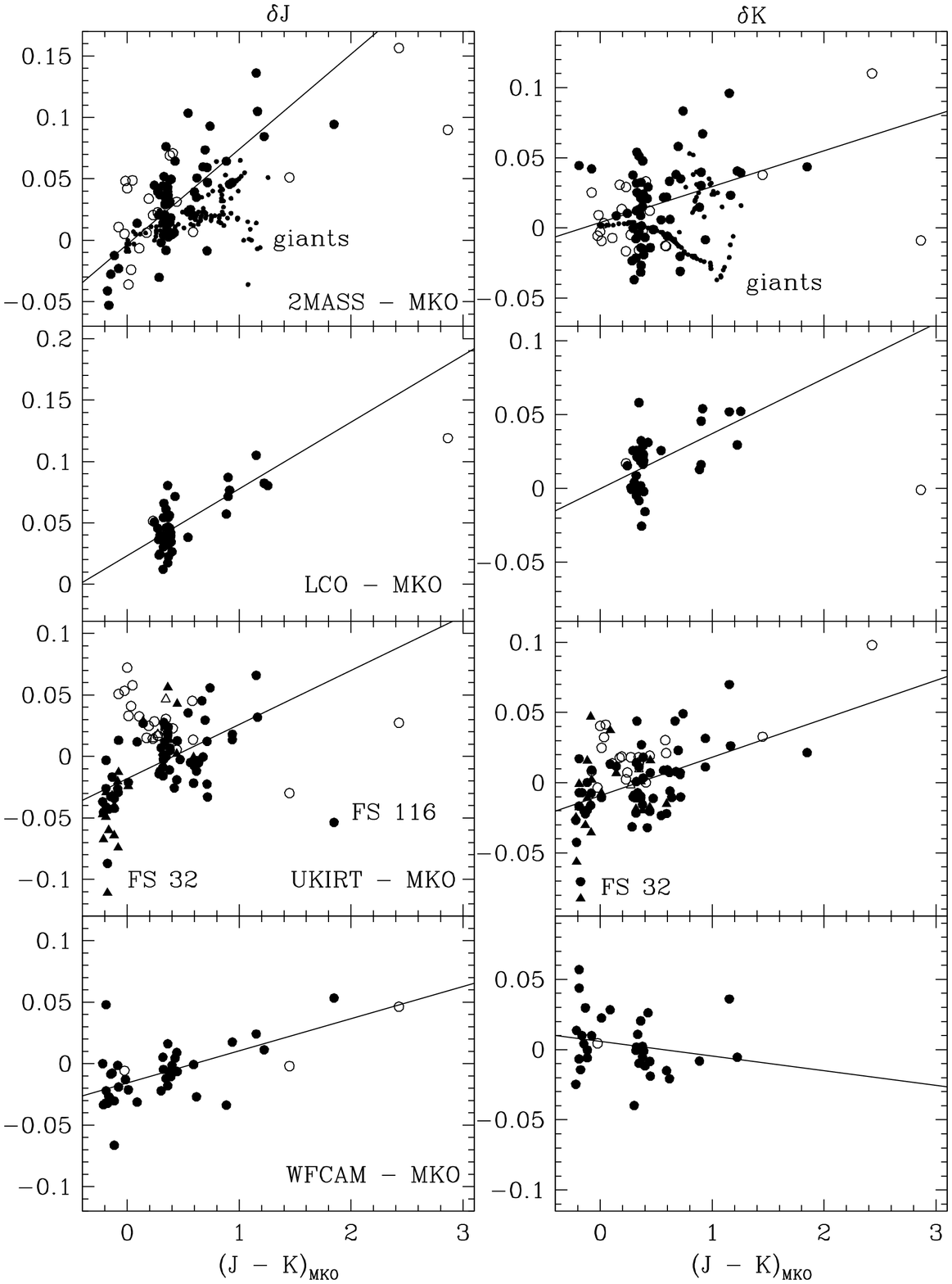}
\caption{Differences between MKO-system $J$ and $K$ magnitudes and those of the 
2MASS, LCO, UKIRT and WFCAM systems, as a function of $J-K$ colour.  Filled circles are
stars with $J>11$ and $K>10.5$,
open are brighter stars that are not included in the
linear fit, shown as a solid line.  Triangles are data on the UKIRT system from
Casali \& Hawarden (1992). Small dots in the 2MASS plots in the top panels are synthetic
results derived by \citet{He} from spectra of dwarf and giant stars; red giants are brighter in 2MASS $J$ and $K_s$  than in MKO $J$ and $K$, and their location is indicated by the ``giants'' label.  FS 32 and FS 116 are identified in the UKIRT plots and are excluded from those fits.  
The MKO, WFCAM and 2MASS $JHK$ and $J$ are fainter than the UKIRT-system values for FS 32 and FS 116, respectively. 
The possible variables FS 144 and GSPC S705-D have been omitted from all plots.
}
\label{JK-fig}
\end{figure}

\begin{figure}
\includegraphics[scale=.45]{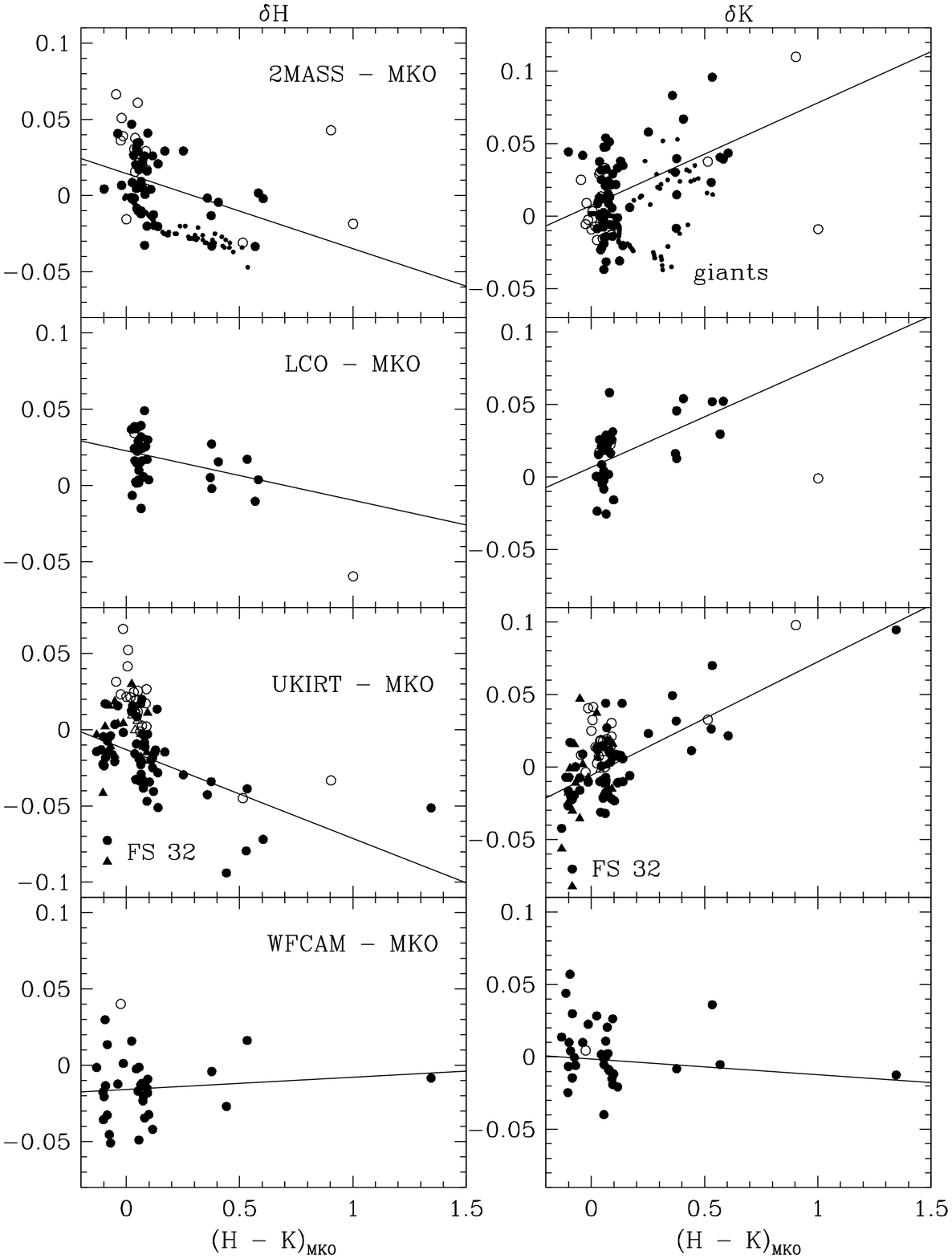}
  \caption{Differences between MKO-system $H$ and $K$ magnitudes and those of the 
2MASS, LCO, UKIRT and WFCAM systems, as a function of $H-K$ colour.  Symbols and labels are as in Figure 2,
except here 
filled circles are stars with $H>10.5$ and $K>10.5$,
open are brighter stars that are not included in the
linear fit, shown as a solid line.
FS 32 is indicated in the UKIRT plot, and excluded from the linear fits. The possible variables FS 144 and 
GSPC S705-D have been omitted from all plots. }
\label{HK-fig}
\end{figure}

The average number of nights each star was observed during this programme is 4.  
One LCO standard star, GSPC S705-D,
appears to be variable, as 4--5 measurements produced standard deviations of 3-6\% across
$J$, $H$ and $K$. \citet{Ps}, however, found a small 1\% deviation over their four measurements
of this star.  The UKIRT standard FS 144 (Ser-EC84) has been reported as possibly variable at the
5\% level, and possibly binary (N. Cross, M. Connelley, private communications), although
our three measurements agree to 3\% at $J$ and 1\% at $H$ and $K$. There are two UKIRT standards
for which the magnitudes presented here are fainter than the UKIRT-system values 
\citep{Ha,Ca} by $\sim$0$\fm$1~: FS 116 (B216-b7) and FS 32 (Feige 108, a subdwarf B star).
The latter is significantly fainter at all $JHK$, the former at $J$ only.
The 2MASS and preliminary WFCAM values agree with the present results to
$<0\fm05$, suggesting that a change may have occurred between the epoch of the Casali \& Hawarden 
and Hawarden et al. observations (1991 and 1994--1998, respectively) and that of the more recent 
2MASS, MKO and WFCAM observations.  We have indicated all four of these stars, GSPC S705-D,
FS 144, FS 116 and FS32, as possibly variable in Table 3.
An L dwarf companion has been found for the UKIRT standard FS 129
(M dwarf LHS 2397a, \citet{Cl}), but this should not lead to detectable variability in the near-infrared.

Omitting S705-D, the average  uncertainty of the magnitudes listed in Table 3 is $0\fm012$ at $J$, 
and $0\fm011$  at $H$ and $K$.   
These errors do not include systematic effects, and we explore these further below, through comparison 
with other published work.

\section{Comparison with Other Data}

\subsection{Datasets}

In this section we compare our measured MKO $JHK$ magnitudes with those measured for stars
in common by: \citet{Ha} and the earlier UKIRT-system measurements by \citet{Ca};
\citet{Hu} using the ARNICA camera with similar filters to the UKIRT-system set;
the LCO (NICMOS) system from \citet{Ps}; the 2MASS atlas;
and preliminary MKO-system data from UKIRT's Wide Field Camera, WFCAM.  As we are
looking for small effects, for the purpose of these comparisons we restrict each dataset 
to values with uncertainty $< 0\fm025$. 

\begin{table*}
 \centering
 \begin{minipage}{100mm}
  \caption{Colour Transformations.}
\begin{tabular}{@{}crrrrc@{}}
  \hline
$\delta$ mag & constant A & error A & slope B  & error B & colour \\
 \hline
$J_{\rm 2MASS} - J_{\rm MKO}$ & $-0.004$ & $0.006$ &$ +0.078$ & $0.010$ & $J-K_{\rm MKO}$ \\
$K_{\rm 2MASS} - K_{\rm MKO}$ & $+0.004$ & $0.007$ & $+0.026$ & $0.011$ & $J-K_{\rm MKO}$ \\
$K_{\rm 2MASS} - K_{\rm MKO}$ & $+0.008$ & $0.004$ & $+0.071$ & $0.020$ & $H-K_{\rm MKO}$ \\
$H_{\rm 2MASS} - H_{\rm MKO}$ & $+0.014$ & $0.004$ & $-0.049$ & $0.018$ & $H-K_{\rm MKO}$ \\
$J_{\rm MKO} - J_{\rm 2MASS}$ & $+0.001$ & $0.006$ & $-0.069$ & $0.010$ & $J-K_{\rm 2MASS}$ \\
$K_{\rm MKO} - K_{\rm 2MASS}$ & $-0.003$ & $0.008$ & $-0.025$ & $0.012$ & $J-K_{\rm 2MASS}$ \\
$K_{\rm MKO} - K_{\rm 2MASS}$ & $-0.006$ & $0.006$ & $-0.065$ & $0.030$ & $H-K_{\rm 2MASS}$ \\
$H_{\rm MKO} - H_{\rm 2MASS}$ & $-0.014$ & $0.004$ & $+0.049$ & $0.022$ & $H-K_{\rm 2MASS}$ \\
$J_{\rm LCO} - J_{\rm MKO}$ & $+0.023$ & $0.005$ & $+0.055$ & $0.009$ & $J-K_{\rm MKO}$ \\
$K_{\rm LCO} - K_{\rm MKO}$ & $-0.001$ & $0.006$ & $+0.037$ & $0.010$ & $J-K_{\rm MKO}$ \\
$K_{\rm LCO} - K_{\rm MKO}$ & $+0.007$ & $0.004$ & $+0.070$ & $0.018$ & $H-K_{\rm MKO}$ \\
$H_{\rm LCO} - H_{\rm MKO}$ & $+0.023$ & $0.003$ & $-0.032$ & $0.015$ & $H-K_{\rm MKO}$ \\
$J_{\rm MKO} - J_{\rm LCO}$ & $-0.019$ & $0.005$ & $-0.058$ & $0.008$ & $J-K_{\rm LCO}$ \\
$K_{\rm MKO} - K_{\rm LCO}$ & $0.003$ & $0.006$ & $-0.038$ & $0.010$ & $J-K_{\rm LCO}$ \\
$K_{\rm MKO} - K_{\rm LCO}$ & $-0.007$ & $0.004$ & $-0.071$ & $0.020$ & $H-K_{\rm LCO}$ \\
$H_{\rm MKO} - H_{\rm LCO}$ & $-0.023$ & $0.003$ & $+0.034$ & $0.017$ & $H-K_{\rm LCO}$ \\
$J_{\rm UKT} - J_{\rm MKO}$ & $-0.018$ & $0.004$ & $+0.044$ & $0.008$ & $J-K_{\rm MKO}$ \\
$K_{\rm UKT} - K_{\rm MKO}$ & $-0.010$ & $0.004$ & $+0.027$ & $0.006$ & $J-K_{\rm MKO}$ \\
$K_{\rm UKT} - K_{\rm MKO}$ & $-0.006$ & $0.003$ & $+0.078$ & $0.010$ & $H-K_{\rm MKO}$ \\
$H_{\rm UKT} - H_{\rm MKO}$ & $-0.013$ & $0.003$ & $-0.058$ & $0.011$ & $H-K_{\rm MKO}$ \\
$J_{\rm MKO} - J_{\rm UKIRT}$ & $+0.017$ & $0.004$ & $-0.044$ & $0.007$ & $J-K_{\rm UKIRT}$ \\
$K_{\rm MKO} - K_{\rm UKIRT}$ & $+0.009$ & $0.004$ & $-0.027$ & $0.006$ & $J-K_{\rm UKIRT}$ \\
$K_{\rm MKO} - K_{\rm UKIRT}$ & $+0.005$ & $0.003$ & $-0.089$ & $0.012$ & $H-K_{\rm UKIRT}$ \\
$H_{\rm MKO} - H_{\rm UKIRT}$ & $+0.014$ & $0.003$ & $+0.064$ & $0.013$ & $H-K_{\rm UKIRT}$ \\
\hline
\end{tabular}
\end{minipage}
\end{table*}

\subsection{Colour Transformations}

Colour transformations must be derived to correct for the different filter profiles (see Figure 1).  
To do this, we plotted the differences between the MKO data presented here and the six datasets listed above (as $\delta J$, $\delta H$ and $\delta K$), as a function of the MKO colours $J - K$ and $H - K$.  In \S 6.3 we show that there are indications of non-linearity effects for stars brighter than
11th magnitude at $J$ and 10$\fm$5 at $H$ and $K$; hence to derive the colour transformations we used only stars fainter than these limits.

Figure 2 shows $\delta J$ and $\delta K$ as a function of 
$J - K$, and Figure 3 shows $\delta H$ and $\delta K$ as a function of 
$H - K$.  The \citet{Hu} data span a small range in colour and are not shown in the figures.
The \citet{Ca} UKIRT dataset did not span a large colour range, and the existing data
showed close agreement with the later \citet{Ha} data, as expected due to their use of the same filters.
These two UKIRT-system datasets are shown together in one row in Figures 2 and 3.
We adopt the same colour transformation for the two datasets, derived from the more accurate \citet{Ha} results.

In the 2MASS rows in Figures 2 and 3 we also show differences derived from synthetic
magnitudes by \citet{He}.  Hewett et al. determined $JHK$ MKO- and 2MASS-system 
magnitudes from the  Bruzual-Persson-Gunn-Stryker Spectral Atlas\footnote{http://www.stsci.edu/hst/observatory/cdbs/bpgs.html}
and additional M dwarf spectra taken from the literature, for a study of the UKIDSS photometric system.
The authors find that colour transformations between 2MASS and MKO are a function of stellar luminosity class, and this can also be seen in Figures 2 and 3.  Red dwarfs and giants with the same $J - K$ or $H - K$ colour can differ in $J$ or $K$ by 0$\fm$05, between the two systems.  Examining the spectra of an M2V and an M3III star, both with $H - K \sim 0.3$, shows that the wider 2MASS $J$ filter
detects the water bands in the dwarf spectrum, and the redder MKO $K$ filter is sensitive to
the strong CO absorption in the red giant (see the filter bandpasses in Figure 1).   The $H$
filters sample similar spectral features, and the offset seen in $\delta H$ in Figure 3 is simply
a difference in zeropoint.  However, it must be borne in mind that for astronomical
objects with very structured spectral energy distributions, such as T dwarfs, magnitudes 
can be system-dependent at the several-tenths of a magnitude level \citep{St}.  

WFCAM on UKIRT contains MKO-system filters and therefore there should be no colour dependency between
the WFCAM and UFTI magnitudes presented here.  Figure 3 shows that  $\delta H$ and $\delta K$ are
consistent with no colour term, but there is a suggestion of a colour dependency at $J$ in Figure 2.  The 
relationship is heavily weighted by the reddest star and the slope is probably spurious.  This will be investigated further as more WFCAM data are obtained and the WFCAM photometric response better understood.

Table 4 gives colour tranformations between MKO and the 2MASS, LCO and UKIRT systems, where
$$ \delta mag = A\pm error_A + (B\pm error_B * colour) $$
Differences in colours can be calculated from the differences in magnitudes. The LCO
transformation should be treated cautiously as it is not well determined (see Figures 2 and 3).  Also,
as described above, a single transformation cannot describe the 2MASS to MKO relationship to
better than $\sim$5\% due to the intrinsic dependency on luminosity class.    

All the photometric systems considered here are  Vega-based, where Vega is defined to have zero 
magnitude in $J$, $H$ and $K$.  The MKO, UKIRT, WFCAM and LCO systems are linked to \citet{El},
who adopt Vega to be zero in all bands.  2MASS used LCO and UKIRT calibrators to calibrate 
their fields.  Thus, the linear fits shown in Figures 2 and 3 should go through (0,0).  
Table 4 (columns 2 and 3) shows that LCO $J$ and $H$ have a small $\sim$0$\fm$02   offset, significant at the
4--7~$\sigma$ level (as also seen in the UKIRT comparison by \citet{Ha});
2MASS $H$ and UKIRT $JHK$ have smaller 0$\fm$010--0$\fm$015 offsets significant at the 3~$\sigma$ level.
These may reflect small drifts introduced into the systems' zeropoints with each step
away from the initial, bright, \citet{El} reference standards (for example the MKO dataset presented 
here is linked to the UKIRT \citet{Ha} dataset which is linked to the \citet{Ca} dataset,
which used \citet{El} standards).

\begin{figure}
\includegraphics[scale=.45]{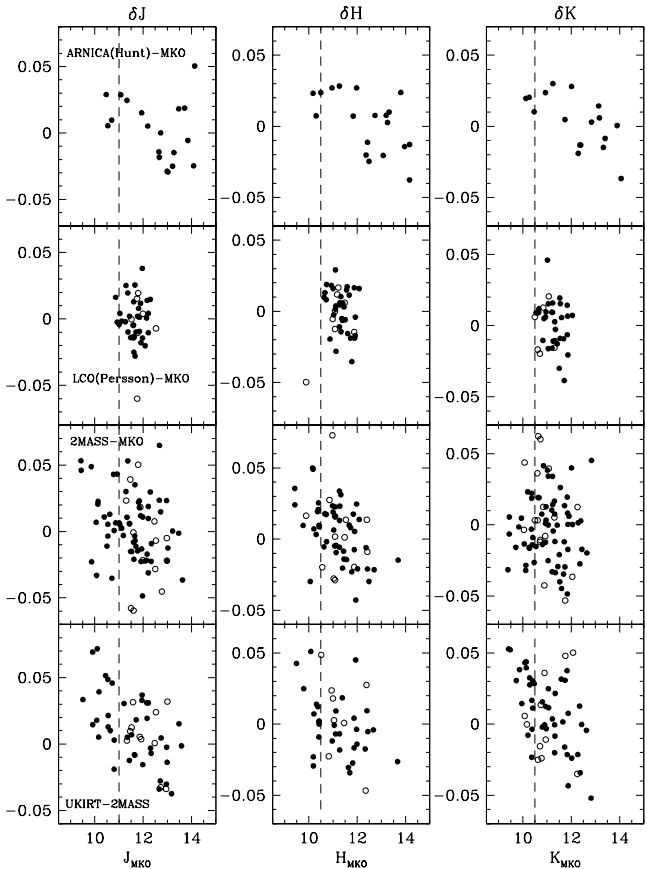}
  \caption{Differences between MKO-system $J$, $H$ and $K$ magnitudes and those of the 
ARNICA, LCO and 2MASS systems, after transformation to MKO,  as a function of $JHK_{\rm MKO}$ brightness.
A comparison between UKIRT \citep{Ha} and 2MASS is also shown, where the 2MASS data has been transformed to the
UKIRT system, and the results plotted against UKIRT $JHK$.  Filled circles are stars
bluer than $J - K =$ 0.7, which have more accurate colour transformations between systems,
and open circles are redder stars. The dashed line separates stars brighter and fainter than
$J=11$ and $H$,$K=10.5$, see discussion in text. The possibly variable stars FS 144, FS 32 and GSPC S705-D
have been omitted from all plots, and FS 116 is excluded in the UKIRT $J$ plot.}
\label{linear-figB}
\end{figure}

\begin{figure}
\includegraphics[scale=.45]{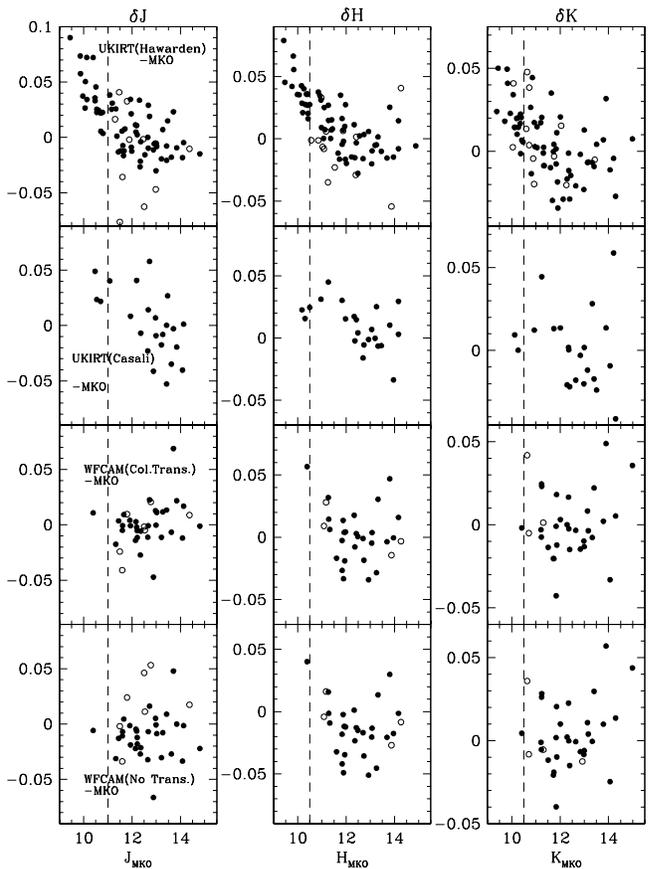}
  \caption{Differences between MKO-system $J$, $H$ and $K$ magnitudes and those of the 
UKIRT and WFCAM systems, as a function of $JHK_{\rm MKO}$ brightness. 
The UKIRT data have been transformed to the MKO system, the WFCAM data are shown both with 
and without a colour transformation. Symbols are as in Figure 4.
UKIRT(Casali) is Casali \& Hawarden (1992) data, and UKIRT(Hawarden) is Hawarden et al.
(2001) data. The possibly variable stars FS 144, FS 32 and GSPC S705-D
have been omitted from all plots, and FS 116 is excluded in the UKIRT $J$ plots.
}
\label{linear-figA}
\end{figure}

\subsection{Data Comparison and Non-Linearity Effects}

\subsubsection{$\delta mag$ as a function of brightness}

Having derived the colour transformations between systems, we can now  
compare our results with other published values. To do this, we have applied the colour transformations
derived above to each dataset, and calculated the difference in magnitudes.  Due to the
fact that our monotonic transformation does not take luminosity class into account,
and hence may introduce an error for redder stars,  the primary
comparison sample consists of stars with $J - K < 0.7$, as well as uncertainties in magnitudes 
$< 0\fm025$.  Red and blue stars are differentiated by symbols in Figures 4 and 5, which
plot $\delta mag$ as a function of $mag_{\rm MKO}$
for the ARNICA, LCO, 2MASS, UKIRT and WFCAM samples.  
The WFCAM comparisons are plotted 
both with the colour terms suggested in Figures 2 and 3 applied, and with no colour term applied.

The clearest trends in Figures 4 and 5 are seen in the UKIRT $-$ 2MASS and UKIRT $-$ MKO comparisons.
The UKIRT $-$ MKO correlation is such that brighter stars are fainter in the UKIRT system or brighter
in the MKO system.
Due to the opposite senses of the non-linearity behaviours of the cameras used for the 
UKIRT and MKO datasets, this could point to a problem with either dataset, or both.  
In fact, the unusual UFTI response is such that
it exaggerates any non-linearity in any of the comparison datasets, if any non-linearity remains in 
the UFTI data.  The comparisons with 2MASS for the UKIRT Hawarden et al. data and the MKO data presented here (bottom two rows in Figure 4) suggest a correlation with the UKIRT data, and a less significant 
correlation with the MKO data (linear regression t-statistics imply the slopes are significant at
the 4, 2 and 5~$\sigma$ level at $J$, $H$ and $K$ respectively for the UKIRT comparison, and 
2, 4 and 0~$\sigma$ for the data presened here). 

These observed trends
spurred an investigation of non-linearity effects in the MKO, UKIRT, WFCAM and 2MASS datasets.
Large scatter or small magnitude range in the samples considered here make the comparison to the \citet{Ca,Hu,Ps} data inconclusive.

\subsubsection{Applied linearity corrections}

The 2MASS data does not have a linearity correction applied as such, but
all raw data are inspected and for any data with signal at the level of 1\% deviation
from linearity, the first read of the image was used as opposed to the two-read
image (R. Cutri private communication).  As the ``Read2 - Read1''  exposure saturation 
threshold is around $K_{\rm s}=$ 8.0, the $K>9$ stars studied here 
should be safely in the linear regime for the 2MASS data, given the option of using 
the first-read data only.  

The linearity response of the WFCAM detectors appears to be flat to 1\% over a large range
in counts (Hodgkin, private communication).  However stars 11th magnitude and brighter will
have peak pixels approaching saturation, and aperture photometry on such stars will be
too faint by $\gtrsim 0\fm02$.  No trends are seen with brightness for the WFCAM data 
shown in Figure 5,
suggesting that, at least for this fainter subset of our sample, the WFCAM and MKO 
data are linear.

We re-reduced the UKIRT and MKO data, taken with the IRCAM and UFTI cameras respectively, 
exploring how changes in the
linearity correction affect the data.  Doubling the correction applied resulted in a
$\leq 0\fm01$ change in the final result.  This is due both to the small
size of the correction, and also to our way of reducing both the IRCAM and UFTI 
data in a relative sense, deriving
an average zeropoint for each night's set of stars.  However problems with the linearity 
correction may still exist.

\subsubsection{Errors in the applied linearity corrections}
 
\citet{Ps} point out that it is easy to underestimate non-linearity effects
for very bright stars, with short exposure times, when using double-read mode.  
Both reads of the pair may be further up the non-linear curve than realised.  

We can estimate the size of this effect for the data presented here and by \citet{Ha}
by assuming that the correction to each individual
read is the same as that derived experimentally for the double-read mode.
If the exposure time is equal to the read time, then the first read will have
counts $\sim$half those of the second read.  In this case, for UFTI, 
if the second read is say 10000~DN, then applying
the linearity correction to each read instead of the difference of the reads,
leads to a signal difference of 6.1\% for this pixel, where the true value is lower.  
If the exposure time is twice
the read time, so that the first read is $\sim$one-third of the second, and the second
read is 9000~DN, then the difference is 3.6\% for that pixel.  The same 
calculations for IRCAM  also produce a $\sim$4\% effect, although here
the true signal level is higher. 

The effect on the
aperture photometry is smaller than this; on a night with UKIRT's typical 0$\farcs$7  
seeing, around one-third of the total flux would arise from pixels with counts
within a factor of two of the maximum.  Hence a 1--2\% effect is likely
for stars observed with exposure times 1--2$\times$ the read time, or 9th--11th
magnitude stars for the MKO (UFTI) and UKIRT (IRCAM) samples, in average to good seeing,
at $J$, $H$ and $K$.  For stars fainter 
than 11th magnitude, the exposure time is $\geq$3--5$\times$ the read time, and
the effect on the aperture photometry is calculated to be $\leq$0.5\%.
This means that the UKIRT to MKO comparisons in Figures 4 and 5 would show
a $\sim$0$\fm$04 effect for bright stars, and the 2MASS comparisons would show a
0$\fm$02 effect (assuming the 2MASS data are linear), as is in fact seen.  

The combination of exposure times and filters used for this work, and by Hawarden et al.,
led to counts on the standard star being larger at $J$ 
than at $H$ and $K$,  for both datasets.
Hence the error in the non-linearity correction
is significant at fainter values of $J$ than of $H$ and $K$.
Linear regression to the UKIRT $-$ MKO plots shown in Figure 5  
gives correlation coefficients and t-statistic values of
0.7 and 8, 0.7 and 9 and 0.5 and 5, for $J$, $H$ and $K$, respectively,
when the entire sample is considered.  
Excluding stars with $J<$11 and $H$, $K <$ 10.5 reduces these to
insignificant values of 0.4 and 2.5, 0.4 and 3 and 0.2 and 1.5.
 
Given the stronger correlation between the 2MASS and UKIRT values than between
2MASS and MKO (Figure 4), and the
larger pixels in IRCAM than UFTI (0$\farcs$28 cf. 0$\farcs$09), 
its likely that more of the non-linearity error lies in the earlier 
\citet{Ha} data than in the data presented here.  In any case, the systematic
effect on the data presented here is small: $+0\fm005$ -- $+0\fm020$
for the 25 stars with $J<$11 or $H$, $K<$ 10.5. For the 90 fainter
stars in the sample, the error is a factor of two, or more, less than the measurement error,
and hence insignificant.

\section{Conclusions}

We have presented $JHK$ magnitudes for 115 stars in the brightness range of 10--15th magnitude.
The photometry is in
the widely adopted MKO photometric system, which uses well-defined filters matched to
the atmospheric windows. 
The internal accuracy is on average 0$\fm$011, with the stars being observed
typically over 4 nights.   The results presented here should be used in preference
to those of \citet{Ha} if observations are being made in the MKO system, as the filters,
and hence the photometric systems, are significantly different from the UKIRT system.

We present colour transformations between the MKO and the 2MASS, LCO and UKIRT photometric
systems.  However these transformations can be luminosity-class dependent at the 5\% level
for redder stars with strong absorption bands.  For very accurate transformations between 
systems, the exact spectral type needs to be known and transformations calculated and applied 
as a function of type (or the stars should be observed in the required system).

We find that the difference between the $JHK$ magnitudes presented here and those presented 
previously by \citet{Ha} shows a dependence on brightness.  We explain this in terms of
an under-estimate of the linearity correction for stars with $J<11$ and $H$,$K<10.5$,
such that the photometry presented here and by Hawarden et al. is affected by $\lesssim$0$\fm$02. 
The effect is likely to be more significant in the results of Hawarden et al. than in the 
present results: this is supported by comparisons with 2MASS data.

There are 84 stars in the sample presented here that have
$11 < J < 15$ and $10.5 < H$,$K < 15$, are not suspected to be variable, and have
magnitudes with an estimated error $\leq 0\fm$027; 79 of these have an error of  $\leq 0\fm$020. 
We recommend that these be adopted as primary standards for the MKO
near-infrared ($JHK$) photometric system.

\section*{Acknowledgments}

The United Kingdom Infrared Telescope is operated by the Joint Astronomy Centre on behalf of the U.K. Particle 
Physics and Astronomy Research Council.  This work would not have been possible without the dedicated
effort of all the UKIRT staff.  We are grateful to A. Tokunaga for pursuing the MKO filters and to
P. Hewett and S. Hodgkin for helpful discussion. 
This research has made use of the SIMBAD database, operated at CDS, Strasbourg, France.
This research has also made use of the NASA/ IPAC Infrared Science Archive, which is operated by the Jet 
Propulsion 
Laboratory, California Institute of Technology, under contract with the National Aeronautics and Space 
Administration. This publication makes use of data products from the Two Micron All Sky Survey, which is a 
joint project of the University of Massachusetts and the Infrared Processing and Analysis Center/California 
Institute of Technology, funded by the National Aeronautics and Space Administration and the National Science 
Foundation. The Starlink Project was funded by the Particle Physics and
Astronomy Research Council and managed by the Space Science
and Technology Department of the Central Laboratory of the
Research Councils. We are grateful to the referee whose comments led to a much improved manuscript.

\bsp

\label{lastpage}

\end{document}